\documentclass[12pt]{iopart}
\usepackage{graphicx}
\usepackage[backend=biber,bibencoding=utf8,
  language=auto,%
  style=numeric-comp,%
  sorting=none,
  maxbibnames=10, 
  natbib=true ]{biblatex}

\bibliography{biblio}
\begin{document}

\title[Coherent Compton scattering from hydrogen and helium atoms]{Coherent Compton scattering from hydrogen and helium atoms}

\author{Irina A. Gnilozub$^1$, Alexander Galstyan$^2$, Yuri V. Popov$^{1,3}$, and Igor P. Volobuev$^1$}
\address{$^1$Skobeltsyn Institute of Nuclear Physics, Lomonosov Moscow State University, Moscow 119991, Russia\\
$^2$Institute of Condensed Matter and Nanosciences, Universit\'e Catholique de Louvain, 2 Chemin du Cyclotron, Box L7.01.07, B-1348 Louvain-la-Neuve, Belgium\\
$^3$Joint Institute for Nuclear Research, Dubna, Moscow region 141980, Russia}
\ead{alexander.galstyan@uclouvain.be}

\vspace{10pt}
\begin{indented}
\item[]September 2018
\end{indented}

\begin{abstract}
We develop an approach to describing  coherent Compton scattering
of photons in the keV energy range from hydrogen and helium atoms
based on a relativistic version of the AA-approximation within the
standard perturbative S-matrix formalism. The resulting formulas
for the cross section take into account the effects of the
electron boundness and correctly reproduce the behavior of the
cross section  at large photon energies, where it goes to Thomson
formula, which coincides with the Klein-Nishina-Tamm formula for
the  cross section in the non-relativistic limit.
\end{abstract}

\section{Introduction}
Compton scattering is (quasi)elastic scattering of photons by
electrons, which   takes place for both  free and bound  electrons
and is well studied in the former case. The cross section of this
process in the case of free electrons is given by the famous
Klein-Nishina-Tamm formula \cite{KN,Tamm}. However, Compton
scattering by bound electrons is much less studied both
experimentally and theoretically.

It is a common knowledge that the interaction of photons with
atoms at energies above their ionization energy and below 1 MeV
goes due to two processes. The first one is photoelectric
emission, which takes place only for bound electrons. At the
energies of the order of the electron binding energy photoelectric
emission is the dominating process with a large  cross section
proportional  to Bohr's radius squared. However, the cross section
of this process rapidly goes down with increasing photon energy
and at larger energies the process of Compton scattering becomes
the dominating one. For light atoms the boundary lies in the keV
energy range.

In the perturbative approach, Compton scattering is described by
the second Born approximation \cite{LL,AB}. In the
non-relativistic description of the process, as well as in the
relativistic one, exact Green's function entering the amplitude of
the process involves the summation over the intermediate electron
eigenstates in the field of a nucleus. This summation is the most
difficult problem in the calculations, which makes it necessary to
develop various approximate summation methods. For example, this
problem is discussed in detail in book \cite{DY}.  For
non-relativistic processes with relatively small  energy
transferred by a photon to the atom, the sum is dominated by the
electron states from the continuum in the Coulomb field of the
nucleus, which are often replaced by plane waves. This is the
so-called non-relativistic AA-approximation.

In the relativistic description a similar approximation  consists
in replacing exact Green's function of electron in the external
Coulomb field by free Green's function of electron. Then, in the
perturbative S-matrix formalism, Compton scattering from bound
electrons can be viewed as photoelectric emission of an off-shell
electron, which then emits a photon. It looks as if the electron
absorbs a part of the photon energy and increases its mass, thus
transforming the process to photoelectric emission of a heavier
electron by a less energetic photon, which leads to an increase of
the cross section. This picture gives reason to expect that the
cross section of Compton scattering from bound electrons should
include a term that behaves similar to the cross section of
photoelectric emission, takes into account the effects of
boundness and falls off with growing energy. However, Compton
scattering can take place also at free electrons, and with the
growth of the photon energy the electrons in the atoms can be
approximately considered to be free. For this reason  the cross
section of Compton scattering from bound electrons in the  limit
of high photon energy should be given by the Klein-Nishina-Tamm
formula somehow modified by the electron wave function. However,
this description is used for all photon energies  in the majority
of papers dealing with Compton scattering by bound electrons
\cite{Pisk1,Pratt,Pisk2}.

Since the ionization energy is different for the electrons of
different shells, Compton scattering by bound electrons is a
coherent process only for the electrons of the same shell. Here we
develop an approach to describing coherent Compton scattering of
photons by the electrons bound in hydrogen and helium atoms, which
have only one electron shell. The generalization to  atoms with a
larger number of electron shells is straightforward and consists
in summing the cross sections of Compton scattering from the
electrons of different shells.

The experimental situation in Compton scattering looks as follows.
Nowadays there is a great demand for coincidental experiments,
where the angles and energies of the outgoing electron and photon
(more precisely, of the ion-residue) are measured simultaneously.
Such experiments allow one to get a deeper insight into the
picture of the momentum distributions in target atoms and,
probably, to expand and to complement the information obtained
from other coincidental experiments, for example, from EIS
(electron impulse approximation) \cite{popov, WM}. However, the
differential cross section measured in such experiments  is 5-7
orders of magnitude lass than the typical cross section of
photoelectric emission, which is a serious problem. Nevertheless,
such experiments are carried out (see, for example,
\cite{Bell,Bell1}) and, to the best of our knowledge, new
experiments are planned.  It is worth noting that such experiments
usually deal with photons in the energy range of several keV.

In the present paper we  consider Compton scattering by electrons
in the ground states of hydrogen and helium atoms  in a
relativistic version of the AA-approximation. The process of
Compton scattering by bound electrons is characterized by thee
parameters of the dimension of energy: the bound state  energy
$\varepsilon<0$,  the photon energy $\omega_1$, and the electron
mass $m$. We will consider the process in the most interesting
energy range of several keV, where the dimensionless quantities
$|\varepsilon|/\omega_1 \ll 1$ and $\omega_1/m \ll 1$ can be
neglected.  However, the ratio of these quantities $m
|\varepsilon|/\omega_1^2$ can be of the order of unity and should be
kept in the formulas. It turns out that the corrections due to
this ratio  are significant and  take into account the electron
boundness.

The approach is based on the standard S-matrix formalism, which
seems to be most suitable for describing this process. This is due
to the fact that the standard perturbative S-matrix formalism
naturally considers this process as photoelectric emission of
off-shell electrons, which then emit a photon and return to the
mass shell. Since the process takes place in the Coulomb field of
the nucleus, it turns out to be convenient to use the S-matrix
formalism in the Furry representation.

Throughout the paper we will follow the conventions of textbook
\cite{LL} and use the Gauss system of units adopted in atomic
physics, as well as the relation $\hbar = c =1.$

\section{Theory}

\subsection{Compton scattering by hydrogen atoms}
In order to make the presentation  clearer, first we will
calculate Compton scattering by electrons in the ground state of
hydrogen atoms in the second order of perturbation theory. Compton
scattering from electrons in helium atoms differs from it only in
two aspects that will be discussed later. The relevant part of the
second order S-matrix in the Furry representation looks like
\begin{equation}\label{S-matrix}
S^{(2)}  =  i\,e^2 N \int \bar \psi(x_2) \hat A(x_2) S_e(x_2,x_1)
\hat A(x_1)  \psi(x_1) dx_1 dx_2,
\end{equation}
where N denotes the normal ordering of the operators, $ \hat A(x)
= \gamma^\mu A_\mu(x) $, and $S_e(x_2,x_1)$ is Green's function of
electron in an external Coulomb field. The field $\psi(x)$ should
also be expanded in the electron eigenstates in this external
field,  and $ \bar \psi(x_2)$ denotes the Dirac conjugate of
$\psi(x_2)$.

\begin{figure}[!h]
\centering
\includegraphics[width=0.75\textwidth]{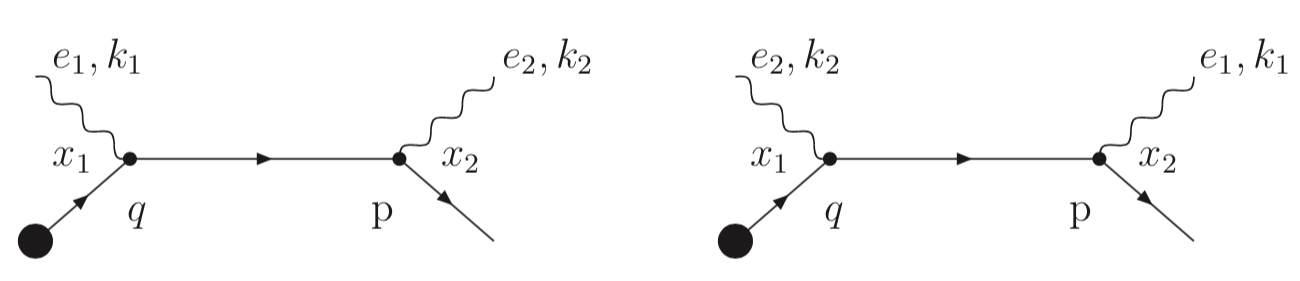}
\caption{Scattering diagrams.}
\label{fig_scattering}
\end{figure}

The processes of scattering, corresponding to this S-matrix term,
are represented by the diagrams depicted in Figure \ref{fig_scattering}.
Here the incoming photon is the one with the polarization and
momentum {$e_1,  k_1$} and the scattered photon is the one with
the polarization and momentum {$e_2, k_2 $}. Below we will
denote the photon energies $k_1^0 = \omega_1$ and $k_2^0 = \omega_2$.

As we have explained in the Introduction, we will use the
following approximation: we will assume that the incoming electron
line marked by the filled circle corresponds to the electron
ground state wave function $e^{-i(m+\varepsilon)t_1} \psi_0(\vec
x_1)$ with the  energy $\varepsilon = -13.62\, $eV in the Coulomb
field of proton, whereas the two other electron lines correspond
to Green's function $S(x_2 - x_1)$  of the free electron field and
to the electron plane wave with momentum $\vec p$. Substituting
the corresponding wave and Green's functions into
(\ref{S-matrix}), integrating with respect to $x_1$ and $x_2$,
denoting $ m + \varepsilon = q^0$ and dropping the delta function
$2\pi \delta (k_1^0 + q^0 - k_2^0 - p^0)$ of energy conservation,
we get the following amplitude of the process:
\begin{eqnarray} \label{amps}
M &=& - 4 \pi e^2 e_2^{*\mu} e_1^\nu \left( \bar u(p) Q_{\mu\nu}
\tilde \psi_0 (\vec q)\right)\\\label{amps_Q} Q_{\mu\nu}&=&
\gamma_\mu \frac{\hat q + \hat k_1 + m}{(q + k_1)^2 - m^2}
\gamma_\nu + \gamma_\nu \frac{\hat q - \hat k_2 + m}{(q - k_2)^2 -
m^2} \gamma_\mu,
\end{eqnarray}
where $\hat q = \gamma_\mu q^\mu,\,\, q = (m + \varepsilon, \vec
q),\,\,  \vec q = \vec p + \vec k_2 - \vec k_1$, $\bar u(p)$ is
the Dirac conjugate of the spinor  $ u(p)$, and the Fourier
transform of the wave function is defined as
\begin{equation}\label{Fourier_trans}
\tilde \psi_0(\vec q) = \int e^{-i\vec q \vec x} \psi_0 (\vec x) d
\vec x.
\end{equation}

This amplitude implies that we have to use a relativistic wave
function of the hydrogen ground state $\psi_0(\vec x)$, which is a
Dirac spinor. However, for our purpose it is sufficient to take
the non-relativistic hydrogen wave function with the first
relativistic correction. Thus, the wave function   $\psi_0(\vec
x)$ will be taken in the form (see \cite{LL}, \S 57)
\begin{equation}\label{wave_function}
\psi_0(\vec x) = \left(1 - \frac{i \gamma^0 \vec \gamma \vec
\nabla}{2 m} \right) \frac{u}{\sqrt{2m}} \psi_{nr}(\vec x),
\end{equation}
where $u$ is a solution of the Dirac equation for the electron at
rest normalized by the condition $\bar u u =  2m,$ and
$\psi_{nr}(\vec x)$ is the non-relativistic wave function of the
ground state of the hydrogen atom. The wave function $\psi_0(\vec
x)$ is normalized to unity up to a term of the order of
$|\varepsilon|/m$, which is negligibly small. Such terms will be
always dropped in our calculations.

Now to find the cross section of the Compton scattering process,
first we have to calculate the squared amplitude. According to the
general rules, the common factor in front of the squared amplitude
will be
\begin{equation}\label{delta_factor}
2\pi \delta (k_1^0 + q^0 - k_2^0 - p^0) .
\end{equation}
Here we will calculate the cross section of the process with
unpolarized particles. To this end, we have to average the squared
amplitude over the polarizations of the incoming particles and  to
sum it over the polarizations of the outgoing particles.  The
summations over the polarizations of the photons and the outgoing
electron are standard. The summation over the polarizations of the
incoming bound electron gives
\begin{eqnarray*}
&&\left(1 + \frac{ \gamma^0 \vec \gamma \vec q}{2 m} \right)
\frac{\sum_s u^s\bar u^s}{{2m}} \left(1 - \frac{ \gamma^0 \vec
\gamma \vec q}{2 m} \right) =  \left(1 + \frac{ \gamma^0 \vec \gamma \vec q}{2 m} \right)
\frac{\gamma^0 + 1}{{2}} \left(1 - \frac{ \gamma^0 \vec \gamma
\vec q}{2 m} \right) \\
& =& \frac{1}{2m} \left((\hat q + m) +
\gamma^0(\frac{\vec q^{\,2}}{4 m} - \varepsilon) - \frac{\vec
q^{\,2}}{4 m} \right).
\end{eqnarray*}
The terms $\frac{\vec q^{\,2}}{m}$ and $|\varepsilon|$ are
negligibly small compared to $m$ and  can be dropped. This means
that the squared amplitude of Compton scattering from  electrons
bound in hydrogen atoms  averaged over the polarizations of the
incoming photon and electron and summed over the polarizations of
the outgoing particles can be written as
\begin{equation}\label{amp^2}
 \overline{{| M|}^{\,2}} = \frac{16\pi^2e^4}{8m} |\tilde \psi_{nr} (\vec
q)|^2 tr\left((\hat q + \hat k_1 - \hat k_2 +m) Q_{\mu\nu}(\hat q
+ m)Q^{\mu\nu}\right),
\end{equation}
and in calculating the numerator of this amplitude we can put
$q^{\,2} = m^2$, i.e. put the incoming electron on the mass shell,
which is equivalent to neglecting the terms $\frac{\vec
q^{\,2}}{m}$ and $|\varepsilon|$ everywhere in the numerator. The
validity of this approximation has been checked by direct
calculations of the squared amplitude, which turned out to be
rather complicated.

It is convenient to represent the trace  in formula (\ref{amp^2})
as \cite{AB}
\begin{equation}\label{trace}
tr\left((\hat q + \hat k_1 - \hat k_2 +m) Q_{\mu\nu}(\hat q +
m)Q^{\mu\nu}\right) = 32 F,
\end{equation}
where $F$ is a scalar function of the momenta $q, k_1, k_2$ and
the electron mass $m$. In the approximation we use this function
is given by the standard expression, which can be found in \S 86
of textbook \cite{LL}:
\begin{eqnarray}\label{factor_F}
F &=& \left(\frac{m^2}{(q + k_1)^2 - m^2} + \frac{m^2}{(q - k_2)^2
- m^2} \right)^2 \\ \nonumber
&+& \left(\frac{m^2}{(q + k_1)^2 -
m^2} + \frac{m^2}{(q - k_2)^2 - m^2} \right) \\ \nonumber
& - &\frac{1}{4}
\left(\frac{(q + k_1)^2 - m^2}{(q - k_2)^2 - m^2} + \frac{(q -
k_2)^2 - m^2}{(q + k_1)^2 - m^2} \right).
\end{eqnarray}

Now we can write the following expression for the differential
cross section of Compton scattering from hydrogen atom
\begin{equation}\label{cross_section}
d\sigma = 2\pi  \delta (k_1^0 + q^0 - k_2^0 - p^0) \frac{32 \pi^2
e^4}{m \omega_1} F|\tilde \psi_{nr} (\vec q)|^2\frac{d \vec
k_2}{(2\pi)^3 2k_2^0}\, \frac{d \vec p}{(2\pi)^3 2p^0},
\end{equation}
where $\vec q =  \vec p + \vec k_2 - \vec k_1$. The fully
differential cross section of the process is obtained by
integrating this expression with respect to $|\vec k_2|$:
\begin{equation}\label{cross_section_fd}
\frac{d\sigma}{d\Omega_{\vec k_2} d\Omega_edE_e} = \frac{r_0^2 m
|\vec p|}{4\pi^3} \frac{\omega_2}{\omega_1} F\ |\tilde \psi_{nr}
(\vec q)|^2.
\end{equation}
Here $r_0 = \frac{e^2}{m}$ is the classical radius of the
electron, $E_e$ is kinetic energy of the scattered electron and
$\omega_2 = \omega_1 + \varepsilon - \frac{\vec p^2}{2m}$ is
defined by energy conservation.

Now we have to calculate the scalar function $F$ in our
approximation. To this end, let us consider energy-momentum
conservation for  the scattering process:
\begin{equation}\label{emom}
q + k_1 - k_2 = p.
\end{equation}
Squaring both sides of the equation gives
\begin{equation}\label{em_conserv}
2m\varepsilon - \vec q^{\,2} + \varepsilon^2 + 2(m +
\varepsilon)(\omega_1 - \omega_2) - 2 \vec q(\vec k_1 - \vec k_2)
- 2\omega_1 \omega_2 + 2\omega_1 \omega_2\cos \theta = 0,
\end{equation}
where $\theta$  is the angle between the momenta $\vec k_1$ and
$\vec k_2$ of the incoming and the outgoing photons, and we have
dropped the term $\varepsilon^2$. As we have already explained in
the Introduction, we will always assume that the conditions
$|\varepsilon|/\omega_1 \ll 1$ and $\omega_1/m \ll 1$ are
fulfilled, which means that the photon energy is in the keV energy
range. Then dividing equation (\ref{em_conserv}) by
$2\omega_1\omega_2$ and neglecting the terms
$|\varepsilon|/\omega_1,\, |\varepsilon|/ \omega_2$ and
$\varepsilon^2/\omega_1\omega_2$, we get
\begin{equation}\label{photon_energies}
\frac{m}{\omega_2}  - \frac{m}{\omega_1} +
\frac{m\varepsilon}{\omega_1\omega_2}  - \frac{\vec
q^{\,2}}{2\omega_1\omega_2}
 - \frac{\vec q(\vec k_1 - \vec k_2)}{\omega_1\omega_2} = 1 -\cos \theta.
\end{equation}
We emphasize that the terms in this equation containing
$\varepsilon$ and $\vec q$ can be of the order of unity  and
cannot be dropped here, as well as  in eq. (\ref{em_conserv}).

Now we need to find the factor $F$ in the approximation we use. To
this end, we have to calculate the denominators in  expression
(\ref{factor_F}) for $F$. The first denominator is
\begin{equation}
(q + k_1)^2 - m^2 \simeq  2m\omega_1 - 2\vec q \vec k_1 +
2m\varepsilon - \vec q^{\,2},
\end{equation}
where we have dropped the term $\varepsilon^2$.  Unlike  the
numerator, in the denominator we cannot neglect the small terms $
\vec q^{\,2}$ and $2m|\varepsilon|$, because the deviation of the
virtual electrons from the mass shell is very small. Then we get
\begin{equation}\label{s-term}
\frac{m^2}{(q + k_1)^2 - m^2} \simeq  \frac{1}{2} \left(
\frac{m}{\omega_1} + \frac{\vec q \vec k_1}{\omega_1^2} +
 \frac{m\varepsilon}{\omega_1^2} +\frac{\vec q^{\,2}}{2\omega_1^2}\right).
\end{equation}
Analogously, we find
\begin{equation}\label{u-term}
\frac{m^2}{(q - k_2)^2 - m^2} \simeq  -\frac{1}{2} \left(
\frac{m}{\omega_2} + \frac{\vec q \vec k_2}{\omega_2^2} +
 \frac{m\varepsilon}{\omega_2^2} -\frac{\vec q^{\,2}}{2\omega_2^2}\right).
\end{equation}
The first terms in the right hand sides of formulas
(\ref{s-term}), (\ref{u-term}) are the standard terms for Compton
scattering from electrons at rest, whereas the other terms are the
corrections due to the electrons being bound in a hydrogen atom.

Thus, the expression in the first bracket in formula
(\ref{factor_F}) turns out to be
\begin{eqnarray}
&& \frac{m^2}{(q + k_1)^2 - m^2} + \frac{m^2}{(q - k_2)^2 - m^2} \simeq \\ \nonumber
&& \frac{1}{2} \left( \frac{m}{\omega_1} - \frac{m}{\omega_2}  -
 \frac{m\varepsilon}{\omega_1^2}  -  \frac{m\varepsilon}{\omega_2^2}
 +\frac{\vec q^{\,2}}{2\omega_1^2} +\frac{\vec q^{\,2}}{2\omega_2^2}
+ \frac{\vec q \vec k_1}{\omega_1^2} - \frac{\vec q \vec k_2}{\omega_2^2}\right).
\end{eqnarray}
Here, with the accuracy up to terms of the order of
$|\varepsilon|/\omega_1$ and $\omega_1/m$,   we can replace
\begin{eqnarray}
\frac{\vec q \vec k_1}{\omega_1^2} - \frac{\vec q \vec k_2}{\omega_2^2}& \rightarrow &
\frac{\vec q(\vec k_1 - \vec k_2)}{\omega_1\omega_2}, \\
-\frac{m\varepsilon}{\omega_2^2} +\frac{\vec q^{\,2}}{2\omega_2^2}&
\rightarrow & -\frac{m\varepsilon}{\omega_1\omega_2} +\frac{\vec
q^{\,2}}{2\omega_1\omega_2}
\end{eqnarray}
and  then use relation (\ref{photon_energies}) for the photon
energies. As a result, we get
\begin{equation}
\frac{m^2}{(q + k_1)^2 - m^2} + \frac{m^2}{(q - k_2)^2 - m^2}
\simeq \frac{1}{2} \left( -\frac{m\varepsilon}{\omega_1^2}
+\frac{\vec q^{\,2}}{2\omega_1^2} - (1 - cos^2 \theta)\right).
\end{equation}

In a similar way one can find that the third term in formula (\ref{factor_F}) is reduced, up to terms of the order of $|\varepsilon|/m$, to
\begin{equation}
 - \frac{1}{4} \left(\frac{(q + k_1)^2 - m^2}{(q - k_2)^2 - m^2} + \frac{(q - k_2)^2 - m^2}{(q + k_1)^2 - m^2}\right) \approx
\frac{1}{4}\left( \frac{\omega_1}{\omega_2} +
\frac{\omega_2}{\omega_1}\right),
\end{equation}
which,  up to terms of the order of $(\varepsilon/\omega_1)^2$, is
equal to $1/2$.

Based on these results, it is not difficult to find an expression
for $F$ in our approximation:
\begin{equation}
F = \frac{1}{4}\left(\left( -\frac{m\varepsilon}{\omega_1^2}
+\frac{\vec q^{\,2}}{2\omega_1^2}\right)^2 + 2 \left(-
\frac{m\varepsilon}{\omega_1^2}  +\frac{\vec
q^{\,2}}{2\omega_1^2}\right) \cos \theta + (1 + \cos^2 \theta)
\right).
\end{equation}
Then the fully differential cross section (FDCS) can be written as
\begin{eqnarray}\label{cross_section_fd_1}
\frac{d^3\sigma}{d\Omega_{\vec k_2} d\Omega_edE_e} = \frac{r_0^2 m
|\vec p|}{2 (2\pi)^3}  \times\nonumber\\
\left(\left(-
\frac{m\varepsilon}{\omega_1^2} +\frac{\vec
q^{\,2}}{2\omega_1^2}\right)^2 + 2 \left(-
\frac{m\varepsilon}{\omega_1^2}  +\frac{\vec
q^{\,2}}{2\omega_1^2}\right) \cos \theta + (1 + \cos^2 \theta)
\right)|\tilde \psi_{nr} (\vec q)|^2,
\end{eqnarray}
where $\vec q = \vec p + \vec k_2 - \vec k_1$ and we have put the
ratio of the photon energies equal to unity in the approximation
we use. We see that the fully differential cross section depends
on the squared wave function of the electron. Thus, measuring this
fully differential cross section can give information on the
momentum distribution of the electrons in hydrogen atom, which is
to be compared with the distribution given by the electron wave
function of the ground state.

It is interesting to find the  single differential cross section (SDCS)
$\frac{d\sigma}{d\Omega_{\vec k_2}}$, which generalizes the
Klein-Nishina-Tamm formula to the case of the scattering from
bound electrons in the energy range under consideration and could
be obtained by integrating  formula (\ref{cross_section_fd_1})
with respect to $\vec p$. However, the straightforward integration
turns out to be inconvenient. To find this differential cross
section we return to formula (\ref{cross_section}) and identically
rewrite it as follows:
\begin{eqnarray}\label{cross_section_1}
d\sigma = \frac{32 \pi^2 e^4}{m \omega_1}\frac{d \vec
k_2}{(2\pi)^3 2k_2^0}\, \frac{d \vec p}{(2\pi)^3 2p^0} \times \nonumber \\
\int
|\tilde \psi_{nr} (\vec q)|^2 \frac{d \vec q}{(2\pi)^3 }
\,(2\pi)^4 \delta^{(4)} (k_1 + q - k_2 - p)F,
\end{eqnarray}
where the delta function is now four-dimensional, i.e. it includes
the delta function of the particle momenta $\delta (\vec k_1 +
\vec q - \vec k_2 - \vec p)$. We see that the differential cross
section looks like the cross section of Compton scattering from an
off-shell electron  with four-momentum $q = (m + \varepsilon, \vec
q)$ averaged over the momenta with the weight $|\tilde
\psi_{nr}(\vec q)|^2/{(2\pi)^3 }$.

 It is convenient to write SDCS in the form
\begin{equation}
\frac{d\sigma}{d\Omega_{\vec k_2}} = \frac{2e^4}{m
\omega_1} \int\frac{d \vec q}{(2\pi)^3}|\tilde \psi_{nr}(\vec q)|^2
\int\int\frac{\omega_2 d\omega_2 d \vec p}{p^0}\delta^{(4)}(q +
k_1 - k_2 - p) F.
\end{equation}
The integration with respect to $\vec p$ is done with the help of
the delta function $\delta(\vec q + \vec k_1 - \vec k_2 - \vec
p)$, which gives $\vec p = \vec q + \vec k_1 - \vec k_2 $. Since
$\vec p^2/m^2 \ll 1$, we can put $p^0 = m$ in the denominator and
get the following expression for the cross section
\begin{equation}
\frac{d\sigma}{d\Omega_{\vec k_2}} =
\frac{2r_0^2}{\omega_1}\int\frac{d \vec q}{(2\pi)^3}|\tilde
\psi_{nr}(\vec q)|^2 \int\omega_2 d\omega_2\delta(\omega_1 +
\varepsilon - \omega_2 - \frac{\vec p^{\,2}}{2m}) F,
\end{equation}
where $r_0 = e^2/m$ is the classical radius of the electron.

The integration with respect to $\omega_2$ is done with the help
of the delta function of energy conservation. It results in
relation (\ref{photon_energies})  for the  photon energies and in
the multiplication of the function $F$ by a factor, which differs
from unity in terms of the order of $\omega_1/m$ and can be
dropped.

It remains to integrate the cross section with respect to $\vec
q$, i.e. to average over the momentum of the bound electron, as
well as to average over the momentum the other relations.
Averaging relation (\ref{photon_energies}) gives
\begin{equation}\label{photon_energies_a}
\frac{m}{\omega_2}  - \frac{m}{\omega_1} +
\frac{2m\varepsilon}{\omega_1\omega_2}  = 1 -\cos \theta,
\end{equation}
or, equivalently,
\begin{equation}\label{photon_energies_ac}
{\omega_2}  = \frac{\omega_1 - 2|\varepsilon|}{1 +
\frac{\omega_1}{m}(1 -\cos \theta)},
\end{equation}
which generalizes the well-known formula for Compton scattering
from free electrons to the case of bound electrons. However,
unlike in the case of free electrons, since the energy of the
scattered photons ${\omega_2}$ is found by averaging formula
(\ref{photon_energies}), it should be viewed as the mean energy of
the scattered photon beam. The broadening of the energy spectrum
of the scattered photons due to the motion of electrons in
hydrogen atoms is given by
$$
\sqrt{\left< \left(\frac{\vec q^{\,2}}{2m}
 + \frac{\vec q(\vec k_1 - \vec k_2)}{m }\right)^2 \right> - \left< \frac{\vec q^{\,2}}{2m}
 + \frac{\vec q(\vec k_1 - \vec k_2)}{m } \right>^2} \simeq
  \sqrt{\varepsilon^2(\beta -1)} = 2|\varepsilon|,
$$
where $\beta = \langle(\vec q^{\,2})^2\rangle/\langle\vec
q^{\,2}\rangle^2 = 5$ is a parameter that depends only on the wave
function of the electrons in hydrogen atom.

Similarly, averaging the factor $F$ gives
\begin{equation}
\overline F = \frac{1}{4}\left((3 + \beta) \left(
\frac{m\varepsilon}{\omega_1^2}\right)^2 - 4 \left(
\frac{m\varepsilon}{\omega_1^2}\right) \cos \theta + (1 +  \cos^2
\theta) \right).
\end{equation}
 Correspondingly, the differential cross section is found to be
\begin{equation}\label{cross_section_f}
\frac{d\sigma}{d\Omega_{\vec k_2}} = \frac{r_0^2}{2}\left( 8\left(
\frac{m\varepsilon}{\omega_1^2}\right)^2 + 4 \left(
\frac{m|\varepsilon|}{\omega_1^2}\right) \cos \theta + (1 + \cos^2
\theta) \right),
\end{equation}
where we have put $\beta = 5.$ The first term in the brackets  can
be brought to the form
$$4{r_0^2}  \left(
\frac{m\varepsilon}{\omega_1^2}\right)^2 = 4{a_0^2}  \left(
\frac{\varepsilon}{\omega_1}\right)^4,
$$
where $a_0 =  1/m e^2$ is Bohr's radius. The formula  is somewhat
similar to the cross section of photoelectric emission due to its
dependence on the ratio of the binding energy to the photon energy
and to the factor  $a_0^2$. This term takes into account the
boundness of the electrons in hydrogen atoms. The term
$\frac{r_0^2}{2}\left(1 + \cos^2 \theta \right)$ corresponds to
Thomson scattering, to which Compton scattering from free
electrons is reduced  in the energy range under consideration, and
the term $2 r_0^2\left( \frac{m|\varepsilon|}{\omega_1^2}\right)\cos
\theta  = 2 r_0 a_0 \left( \frac{\varepsilon}{\omega_1}\right)^2
\cos \theta $ can be considered as an interference term.

It is necessary to note that the description of the outgoing
electron in the Coulomb field by the plane wave can fail at small
momenta of the electron. This is not critical for the differential
cross section given  by  formula (\ref{cross_section_f}), where we
have integrated over this energy range. However, in formula
(\ref{cross_section_fd_1}) for the fully differential cross
section it can be of importance.

This flaw can be amended as follows. We observe that the squared
absolute value of the Fourier transform of the electron ground
state wave function enters the expression for the fully
differential cross section as a factor. The Fourier transform can
be viewed as the matrix element
$$
\tilde \psi_{nr}(\vec q) = \int \psi_{\vec p}^*(\vec x) e^{i\vec Q
\vec x} \psi_{nr}(\vec x) d \vec x,
$$
where the transferred momentum $\vec Q=\vec k_1-\vec k_2$ and the
initial and the final electron states are
$$
\psi_{nr}(\vec x)=\sqrt\frac{(m e^2)^3}{\pi}\ e^{- m e^2|\vec x|},
\quad \psi_{\vec p}(\vec x) = e^{i\vec p \vec x}.
$$
Thus, to take into account the Coulomb field, we can replace the
plane wave $\psi_{\vec p}(\vec x)$ of the escaped electron by the
Coulomb wave function $\psi_{\vec p}^{(-)}(\vec x)$ in this matrix
element without loss of generality of the previous exposition.
Omitting the details of bulky calculations we obtain
\begin{eqnarray}\label{Coulomb}
\left|\int {\psi_{\vec p}^{(-)}}^*(\vec x) e^{i\vec Q \vec x}
\psi_{nr}(\vec x) d \vec x\right|^2 = \nonumber\\
512\pi^2\frac{(m e^2)^6
Q^2}{ p}\left(\frac{\exp\left[-\frac{2 m e^2}{p}\arctan\frac{2 m
e^2 p}{Q^2-p^2+(m e^2)^2}\right]}{1-e^{-2\pi m e^2/p}}\right)\times
\\ \nonumber
\frac{(Q-p\cos\chi)^2+ (m e^2)^2\cos^2\chi}{[(Q^2-p^2+(m
e^2)^2)^2+4p^2(m e^2)^2][Q^2+p^2+(m e^2)^2-2Qp\cos\chi]^4},
\end{eqnarray}
where $Q = |\vec Q|, \,\, p = |\vec p|$ and $\chi$ denotes the
angle between the vectors $\vec Q$ and $\vec p$. If we formally
take the limit $\ m e^2/p \rightarrow 0$, we reproduce the
explicit expression for $|\tilde \psi_{nr}(\vec q)|^2,$
$$
|\tilde \psi_{nr}(\vec q)|^2=\frac{64\pi (m e^2)^5}{[(\vec Q-\vec
p)^2+(m e^2)^2]^4},
$$
as it should be.

The expression in eq. (\ref{Coulomb}) should be substituted in
formula (\ref{cross_section_fd_1}) instead of the factor $|\tilde
\psi_{nr}(\vec q)|^2.$

\subsection{Compton scattering by helium atoms}

The approach under consideration can be applied to helium atoms
only if we assume that the electrons  are independent and
described by a wave function of Hartree-Fock type. To be specific,
we will describe  the electrons in the ground state of the helium
atom by the non-relativistic hydrogen wave function with the
effective charge $Z^* = 27/16$ and the relativistic correction
defined in eq. (\ref{wave_function}). Although this wave function
is known to give a rather rough approximation for the binding
energy, it has the advantage that all the calculations can be
carried out analytically. Since the corrections to the cross
sections depend on the ratio of the binding energy to the energy
of the  incoming photon, this deviation  changes  the final
results very little.

 We will denote the ground state  energy of a  single electron
in helium atom for our choice of the wave function by
$\varepsilon_s\approx -38.8\, eV$ and its wave function by
$e^{-i(m+\varepsilon_s)t_1} {\psi}_0(\vec x_1)$, and the ground
state energy of the electron in the helium ion by
$\varepsilon_{hi}= -54.4\, eV$ and its wave function by
$e^{-i(m+\varepsilon_{hi})t_1} {\psi}_{i0}(\vec x_1)$. The
ionization potential of a single electron  is  $|\varepsilon_i| =
-2 \varepsilon_s  +\varepsilon_{hi} \approx 23.2\,  $ eV, which is
somewhat less the the experimental value.

It turns out that the Furry representation is not convenient for
describing Compton scattering from helium atoms because it is not
capable of taking into account the electron rearrangement in this
process, which leads to the change of the background Coulomb
field. However, we can use the developed description in terms of
free Green's function for the active electron  and take into
account the second electron that remains in the helium ion by the
standard quantum-mechanical recipe. First, this means that we have
to take into account the electron rearrangement in the energy
conservation equation, which reads
\begin{equation}\label{energy_conservation_H}
\omega_1 +  2(m + \varepsilon_s)  - \omega_2 -(m +
\varepsilon_{hi}) - p^0  = \omega_1 +  (m - |\varepsilon_{i}|)  -
\omega_2 - p^0  = 0.
\end{equation}
Therefore, the energy of the active electron, i.e. the one to be
emitted, should be put equal to $q^0 = m - |\varepsilon_{i}|$,  and
the delta function of energy conservation is $2\pi \delta(k_1^0 +
q^0 - k_2^0 - p^0).$ Second,  amplitude (\ref{amps}) should be
multiplied by the overlap integral
$$
\int {\bar \psi}_0 (\vec x) {\psi}_{i0} (\vec x) d\vec x,
$$
where ${\bar \psi}_0 (\vec x)$ is the Dirac conjugate of ${
\psi}_0 (\vec x)$. Thus, the amplitude of the process of Compton
scattering from helium atom can be written as
\begin{equation} \label{amp_H}
M = - 4 \pi e^2 e_2^{*\mu} e_1^\nu \left( \bar u(p) Q_{\mu\nu}
\tilde \psi_0 (\vec q)\right) \int {\bar \psi}_0^\prime(\vec x)
{\psi}_{i0} (\vec x) d\vec x,
\end{equation}
where $q = (m - |\varepsilon_{i}|, \vec q), \, \vec q = \vec p +
\vec k_2 - \vec k_1$, $Q_{\mu\nu}$ and the Fourier transform of
the wave function are still given by (\ref{amps_Q}),
(\ref{Fourier_trans}), and the wave functions $\tilde \psi_0 (\vec
q)$ and  $ {\psi}_0^\prime (\vec x)$ correspond to opposite values
of the spin projections.

Now we have to calculate the squared amplitude averaged over the
polarizations of the incoming particles and summed over the
polarizations of the outgoing particles. Taking into account that
the ground state electrons in a helium atom have opposite spin
projections, we get
\begin{equation}\label{amp^2_H}
 \overline{{| M|}^{\,2}} = \frac{128\pi^2e^4}{m} F |\tilde \psi_{0} (\vec
q)|^2  C,
\end{equation}
where
$$
C = \left|\int  {\bar\psi}_{0} (\vec x) {\psi}_{i0} (\vec x) d\vec
x\right|^2=\frac{512Z^{*3}}{(Z^{*2}+2)^6}
$$
has been calculated with the non-relativistic wave functions
entering formula (\ref{wave_function}) and neglecting the terms of
the order $\varepsilon/m$, and $F$ is given by (\ref{factor_F}).
This expression for $F$ and energy-momentum conservation equation
(\ref{emom}) are the same as in the case of hydrogen atom with the
replacement of the ionization potential of hydrogen atom
$|\varepsilon|$  by the single ionization potental of helium atom
$|\varepsilon_i|.$ Thus, we can repeat all the calculations of the
previous subsection and obtain the following expression for $F$ in
the case of helium atom
\begin{equation}
F = \frac{1}{4}\left(\left( \frac{m|\varepsilon_i|}{\omega_1^2}
+\frac{\vec q^{\,2}}{2\omega_1^2}\right)^2 + 2 \left(
\frac{m|\varepsilon_i|}{\omega_1^2}  +\frac{\vec
q^{\,2}}{2\omega_1^2}\right) \cos \theta + (1 + \cos^2 \theta)
\right).
\end{equation}
Then the fully differential cross section of Compton scattering
from helium atom is given by
\begin{eqnarray}\label{cross_section_fd_H}
\frac{d^3\sigma}{d\Omega_{\vec k_2} d\Omega_edE_e} = C\frac{r_0^2 m
|\vec p|}{ (2\pi)^3} \times\nonumber\\
 \left(\left(\frac{m|\varepsilon_i|}{\omega_1^2} +\frac{\vec
q^{\,2}}{2\omega_1^2}\right)^2 + 2 \left(
\frac{m|\varepsilon_i|}{\omega_1^2}  +\frac{\vec
q^{\,2}}{2\omega_1^2}\right) \cos \theta + (1 + \cos^2 \theta)
\right)|\tilde \psi_{0} (\vec q)|^2,
\end{eqnarray}
which differs from the case of hydrogen atom in the factor $2C$.

The  single  differential cross section can be found by averaging
over $\vec q$ the cross section of Compton scattering  of an
electron with  4-momentum $q = (m - |\varepsilon_i|, \vec q) $ in
the same way, as it was done for the hydrogen atom:
\begin{eqnarray}\label{cross_section_fH}
\frac{d\sigma}{d\Omega_{\vec k_2}} = C {r_0^2}\left( \left(
\frac{m(|\varepsilon_s| + |\varepsilon_i|)}{\omega_1^2}\right)^2 +\right.\nonumber\\
\left.(\beta - 1) \left( \frac{m\varepsilon_s}{\omega_1^2}\right)^2 + 2
\left( \frac{m(|\varepsilon_s| +
|\varepsilon_i|)}{\omega_1^2}\right) \cos \theta + (1 + \cos^2
\theta) \right),
\end{eqnarray}
where $\beta = 5$ for our choice of the ground state wave
function. The relation between the energies of the incoming and
the outgoing photons is now  given by
\begin{equation}\label{photon_energies_aH}
\frac{m}{\omega_2}  - \frac{m}{\omega_1} - \frac{m(|\varepsilon_s|
+ |\varepsilon_i|)}{\omega_1\omega_2}  = 1 -\cos \theta,
\end{equation}
or, equivalently,
\begin{equation}\label{photon_energies_ac_h}
{\omega_2}  = \frac{\omega_1 - (|\varepsilon_s| +
|\varepsilon_i|)}{1 + \frac{\omega_1}{m}(1 -\cos \theta)}.
\end{equation}
The broadening of the energy spectrum of the scattered photons due
to the motion of electrons in helium atoms is given by $\sqrt
{\varepsilon_s^2 (\beta - 1)}= 2 |\varepsilon_s|$ for our choice of
the helium wave function.

Similar to the case of hydrogen atom, formula
(\ref{cross_section_fd_H}) can be improved by replacing
$$
\tilde \psi_{0} (\vec q) \rightarrow \int {\psi_{\vec
p}^{(-)}}^*(\vec x) e^{i\vec Q \vec x} \psi_{0}(\vec x) d \vec x,
$$
where the function $\psi_{0}(\vec x)$ corresponds to $Z^* = 27/15$
and the function $\psi_{\vec p}^{(-)}(\vec x)$ corresponds to $Z =
1.$   As a rule, the further standard orthogonalization of these
functions is useful.

\section{Results and discussion}
\begin{figure}
\centering
\includegraphics[scale=0.6]{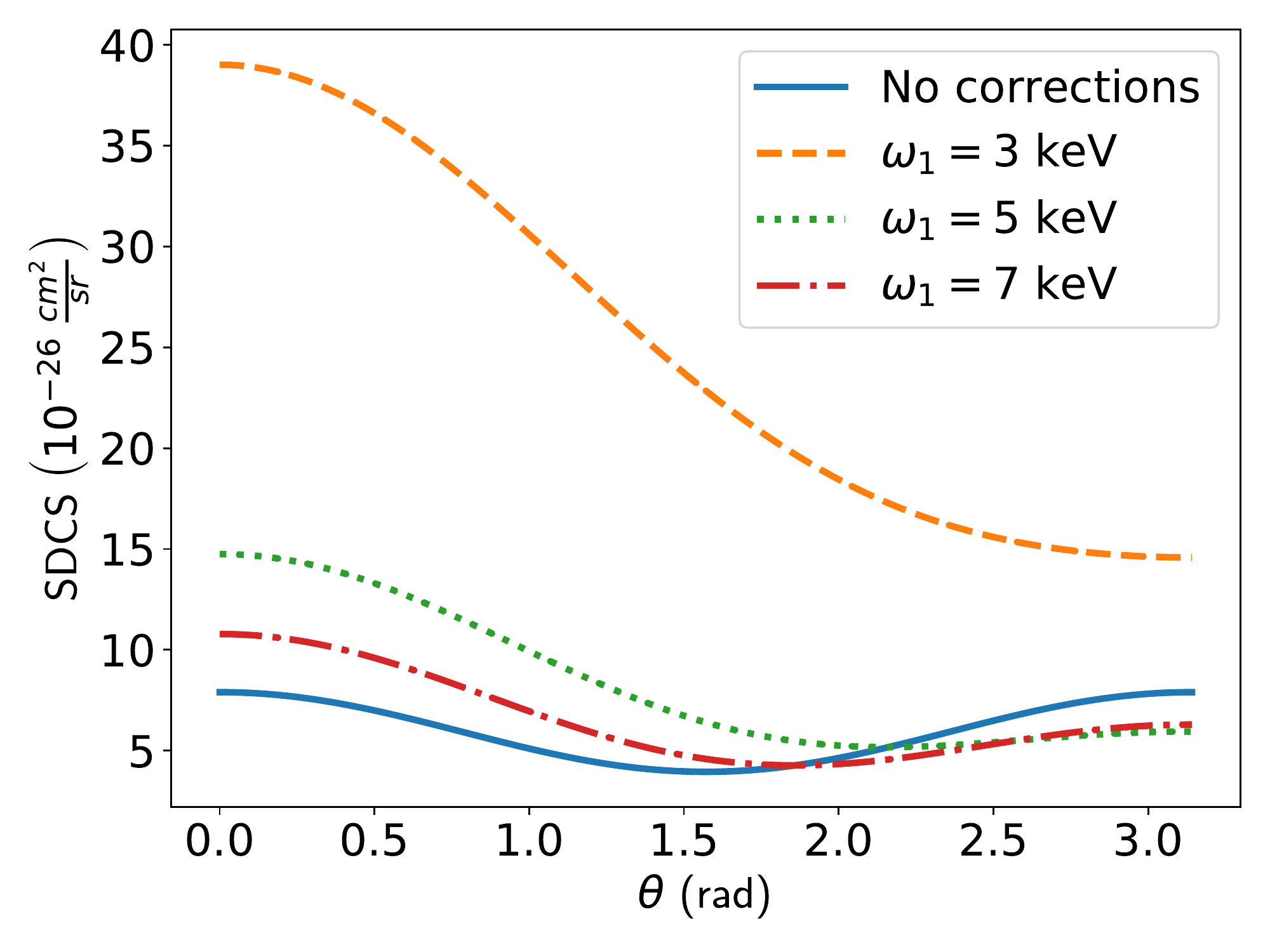}
\caption{\label{fig1} (Color online) SDCS in the case of hydrogen
atom eq. (\ref{cross_section_f}) for three photon energies: 3 keV
(dashed orange line), 5 keV (dotted green line) and 7 keV
(dash-dotted red line). Blue solid line is the Thomson limit.}
\end{figure}

In this section we will present the results of cross section
calculations. We start with the single differential cross section
for the hydrogen atom, which is given by formula
(\ref{cross_section_f}). It is  worth noting  once again that
formulas (\ref{photon_energies_ac}), (\ref{cross_section_f}) are
valid only in the energy range, where  the conditions
$|\epsilon|/\omega_1 \ll 1$ and $\omega_1/m \ll 1$ are fulfilled,
the actual accuracy of the formulas being defined by the larger of
the two ratios. Thus, for Compton scattering from hydrogen, the
best accuracy of about 1\% is expected to be  achieved for the
photon energy $\omega_ 1 \simeq \sqrt{m|\epsilon|} $, which is
approximately 2.64 keV in this case. In Fig. \ref{fig1} the single
differential cross section for the hydrogen atom for the photon
energies 3, 5 and 7 keV is presented, as well as the Thomson
differential cross section for the electron. The accuracy of
formula (\ref{cross_section_f}) in this energy range is expected
to be about 2\%.

We see that the single differential cross section for the hydrogen
atom at 3 keV is essentially larger than the Thomson cross section
and behaves rather differently from it. However, this cross
section  falls rapidly with the growth of the photon energy. The
single differential cross section at 5 keV has a minimum, which
approaches the minimum of the Thomson cross section as the photon
energy goes to 7 keV.   At the photon energy of about 25 keV the
inaccuracy of formula (\ref{cross_section_f})  for the hydrogen
atom becomes larger than its deviation from the Thomson cross
section, and in the energy range above it the Klein-Nishina-Tamm
formula for Compton scattering at free electron can be used to
describe Compton scattering from hydrogen atoms with a high
accuracy.

\begin{figure}
\centering
\includegraphics[scale=0.6]{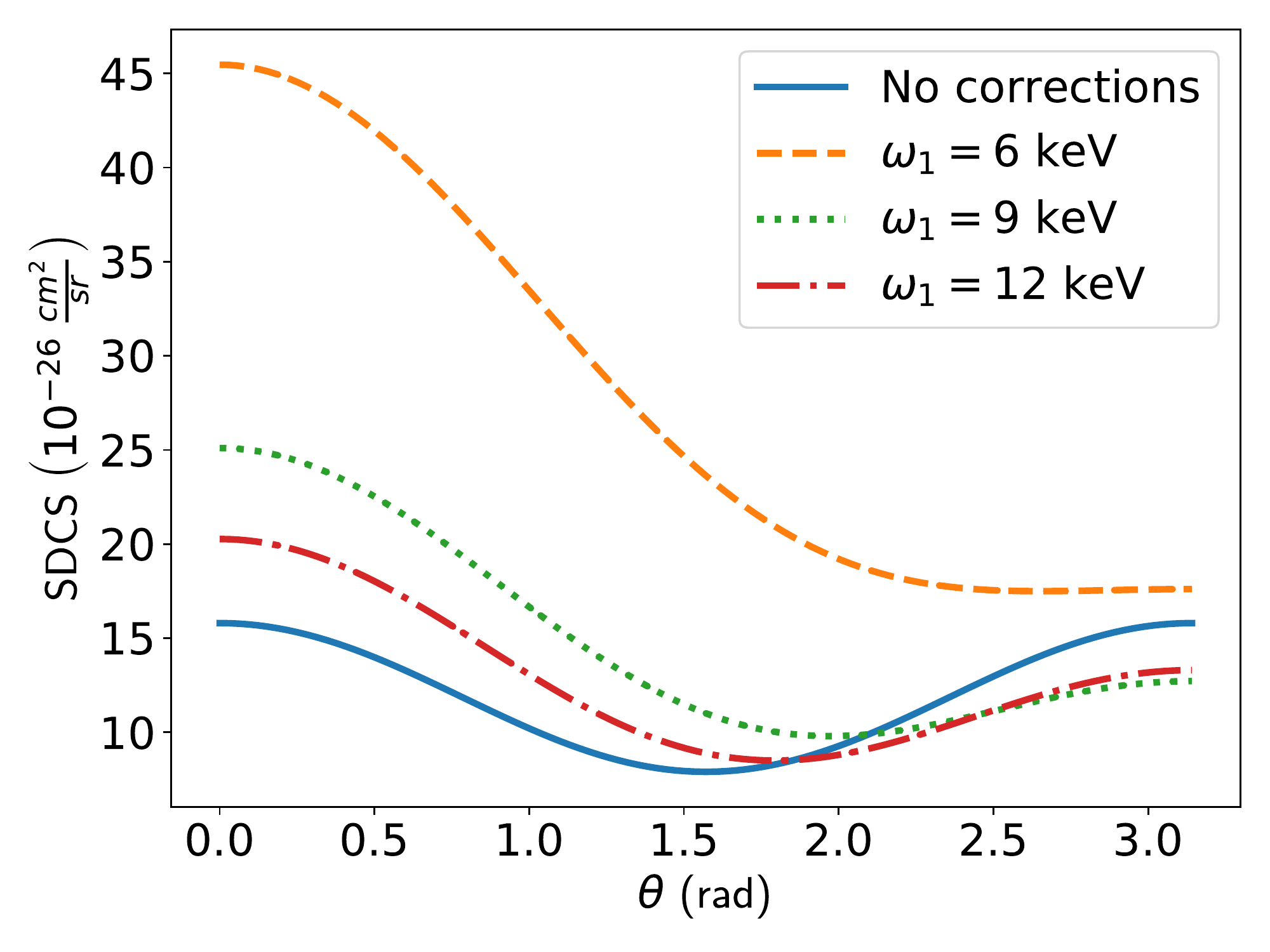}
\caption{\label{fig2} (Color online) SDCS in the case of helium
atom eq. (\ref{cross_section_fH}) for three photon energies: 6 keV
(dashed orange line), 9 keV (dotted green line) and 12 keV
(dash-dotted red line). Blue solid line is the Thomson limit.}
\end{figure}

 Compton scattering from  helium atoms is
described by formula (\ref{cross_section_fH}) and is very similar
to that of hydrogen. The best accuracy of this formula is expected
to be achieved for the photon energy $\omega_ 1 \simeq
\sqrt{m|\epsilon_s|}\simeq 4.45$ keV and is also   about 1\%. The
single differential cross section for the helium atom for the
photon energies 6, 9 and 12 keV, together with the doubled Thomson
differential cross section for the electron, is presented in Fig.
\ref{fig2}.

Again we see that the single differential cross section at 6 keV
is much larger than the Thomson cross section and has a rather
different shape. We also see that, with the growth of the photon
energy, the single differential cross section develops a minimum
and becomes more similar to the Thomson cross section.   At the
photon energy of about 32 keV the inaccuracy of formula
(\ref{cross_section_fH}) for the helium atom becomes larger than
its deviation from the doubled  Thomson cross section, which means
that in the energy range above it the Klein-Nishina-Tamm formula
for Compton scattering at free electrons can be used to describe
Compton scattering from helium atoms.

 Next we pass to calculating the fully differential cross
section of Compton scattering from hydrogen. To calculate this
cross section it is convenient to introduce dimensionless
variables in accordance with the following notations:
$$
\gamma=a_0\omega_1=\frac{\omega_1}{me^2}=\frac{1}{\alpha}\left(\frac{\omega}{m}\right)=0.27\
\omega(keV), \quad \mu=a_0Q \approx \gamma\sqrt{2(1-\cos\theta)},
$$
$$
\quad x=a_0p = \frac{1}{\alpha}\sqrt{\frac{2E_e}{m}}\approx
0.27\sqrt{E_e(eV)}, \quad e^2=\alpha=\frac{1}{137}.
$$

Substituting Coulomb correction (\ref{Coulomb}) into  fully
differential cross section (FDCS) (\ref{cross_section_fd_1}) we get the
following expression for the FDCS in
terms of the dimensionless variables
\begin{eqnarray}
\label{cross_section_fd_2}
\frac{d^3\sigma}{d\Omega_{\vec k_2} d\Omega_edE_e} =
\frac{32}{\pi}(\alpha a_0)^3\mu^2
\left(\frac{\exp\left[-\frac{2}{x}\arctan\frac{x}{\mu^2-x^2+1}\right]}{1-e^{-2\pi
/x}}\right)
\nonumber\\ \times \left[ \left(\frac{1+\mu^2+x^2-2\mu
x\cos\chi}{2\gamma^2} + \cos\theta\right)^2 + 1 \right]\nonumber\\
\times\frac{(\mu-x\cos\chi)^2+
\cos^2\chi}{[(\mu^2-x^2+1)^2+4x^2][\mu^2+x^2+1-2\mu
x\cos\chi]^4}.
\end{eqnarray}
Next we observe that $(\alpha a_0)^3=(1/m)^2(1/m)=1.491\cdot
10^{-21}(cm^2)\times (1/5.11\cdot 10^{5} (eV))=0.29\cdot
10^{-26}(cm^2/eV)$. Thus, we obtain the cross section in
$cm^2/sr^2\cdot eV$.

Integrating formula (\ref{cross_section_fd_2}) with respect to $\chi$
gives us double differential cross section (DDCS), or the energy spectrum of electrons in the case, where the
vectors $\vec k_1, \vec k_2, \vec p$\,\, lie in the same plane (in
the  scattering plane)
\begin{equation}\label{cross_section_fd_3}
\frac{d^2\sigma}{d\Omega_{\vec k_2} dE_e} =(2\pi)\int_0^\pi\
\frac{d^3\sigma}{d\Omega_{\vec k_2} d\Omega_edE_e}\ \sin\chi\
d\chi.
\end{equation}

\begin{figure}
\centering
\includegraphics[scale=0.6]{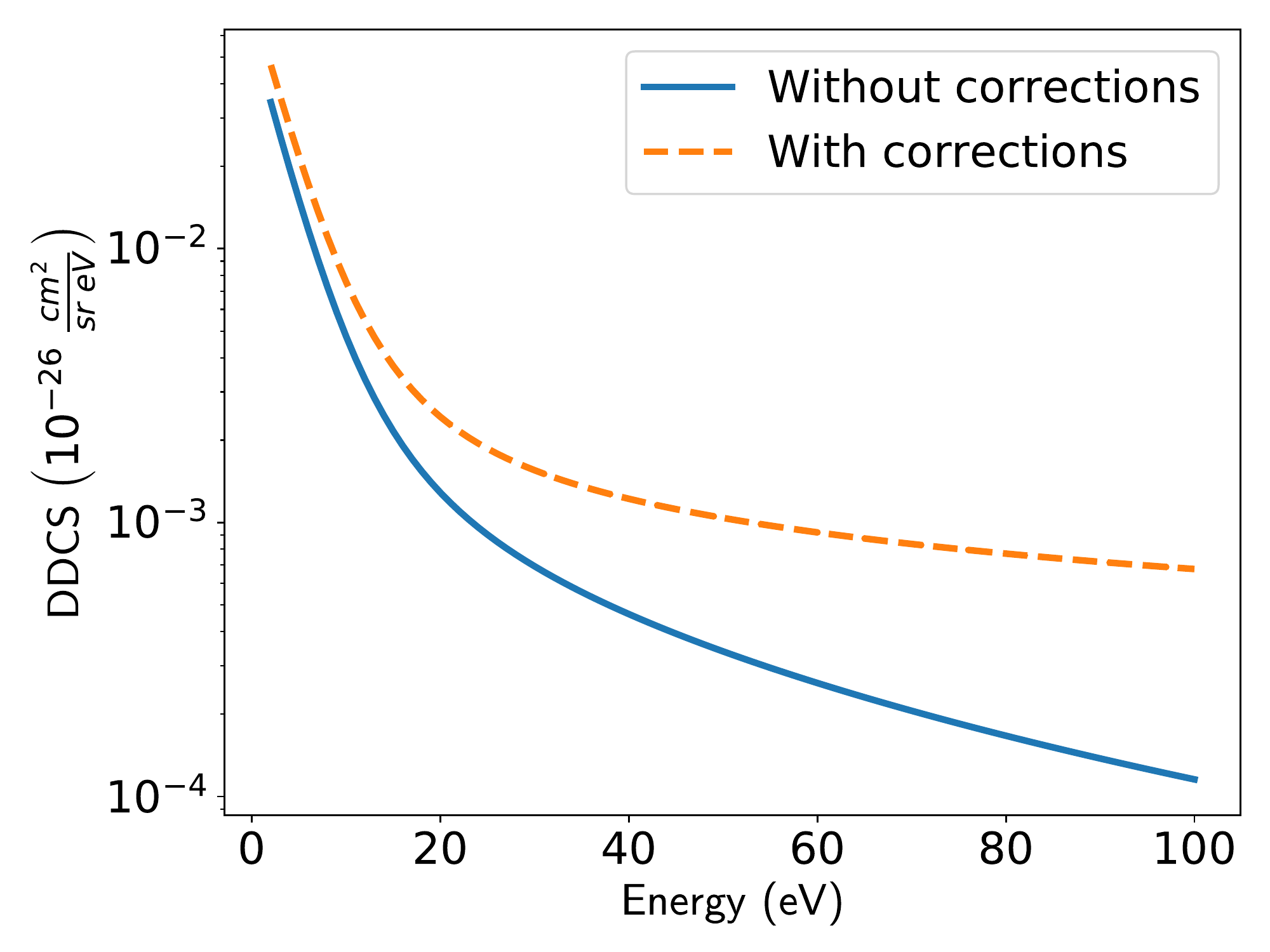}
\caption{\label{fig31} (Color online) DDCS eq.
(\ref{cross_section_fd_3})  versus the energy of the escaped
electron in the case of hydrogen atom. The orange dashed curve
corresponds to eq. (\ref{cross_section_fd_2}) with the correction
terms in the big square brackets. The solid blue line corresponds
to eq. (\ref{cross_section_fd_2}) without these terms. The photon
energy $\omega_1=5$ keV, the scattering angle $\theta=0.1$ rad.}
\end{figure}

\begin{figure}
\centering
\includegraphics[scale=0.6]{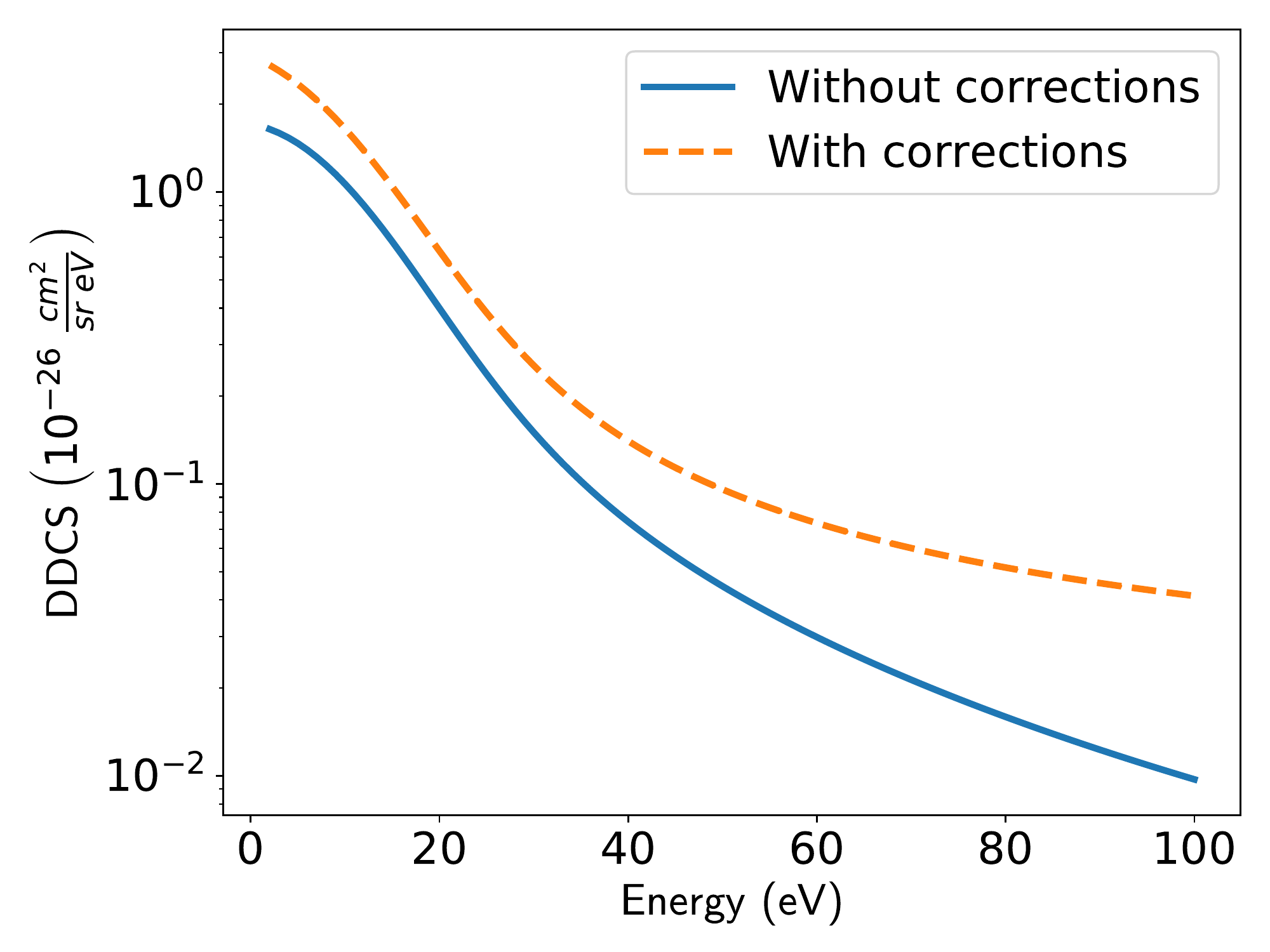}
\caption{\label{fig32} (Color online) The same as in Fig.
\ref{fig31} for the scattering angle $\theta=\pi/4$.}
\end{figure}

 The results of calculating the DDCS with the help of this
formula are presented in Fig. \ref{fig31} for the scattering angle
$\theta = 0.1$ rad and in Fig. \ref{fig32} for the scattering
angle $\theta=\pi/4$.


 We see that the higher the electron energy is, the larger is
the difference between the  curves both in Fig. \ref{fig31} and
Fig. \ref{fig32}. However, in this energy range the DDCS is
extremely small (we note that the DDCS is presented in the
logarithmic scale).

For small electron energies the difference between the curves is
not so noticeable. Besides, we see that larger momentum transfers
$Q$ (big angles $\theta<\pi/2$) increase the absolute value of
DDCS. Thus, we can conclude that, at relatively large momentum
transfers,  Compton scattering from hydrogen is  accessible for
experimental studies. Figs. \ref{fig1} and \ref{fig2} suggest that
the DDCS of Compton scattering from helium atoms should be even
larger.  A detailed study of this case will be carried out in a
separate paper.

\section{Conclusions}

In the present paper we have put forward a new approach to
describing Compton scattering by bound electrons and applied it to
the hydrogen and helium atoms. The approach is based on a
relativistic version of the AA-approximation in the standard
perturbative S-matrix formalism and allows one to describe this
process consistently in the range of the photon energy $\omega_1$
satisfying the conditions $|\varepsilon|/\omega_1 \ll 1$ and
$\omega_1/m \ll 1$. For the hydrogen and helium atoms this is the
energy range of several keV.

The obtained formulas for the cross section take into account the
effects of boundness and correctly reproduce the high photon energy
behavior, which is just the Klein-Nishina-Tamm cross section.

In the present paper we have considered  only the case of
unpolarized photons. However, the results can be easily
generalized to the case of polarized incoming and outgoing
photons. In the case of helium, the formulas can  also be improved
by taking realistic wave functions for the helium ground state.
However, this will result in much more complicated calculations
and will be discussed separately.

\vskip 12pt

\section{Acknowledgements}

The authors thank Prof. R. D\"orner and Dr. M. Sch\"offler for
presenting some materials, which motivated us to carry out this
investigation. We also thank Prof. O. Chuluunbaatar for some help in numerical calculations. Y.P. is grateful to the Russian Foundation for Basic Research (RFBR) for financial support under Grant 16-02-00049-a.

\printbibliography
\end{document}